\documentstyle[aps,prd,preprint]{revtex}   
\begin{document}
\draft
\tightenlines
\newcommand{\half}{{\scriptstyle{\frac{1}{2}}}}
\newcommand{\LP}{\lambda \Phi^4}
\newcommand{\kk}{{\bf k}}
\newcommand{\pp}{{\bf p}}
\newcommand{\qq}{{\bf q}}
\newcommand{\xx}{{\bf x}}

\newcommand{\BE}{\begin{equation}}
\newcommand{\EE}{\end{equation}}
\newcommand{\BA}{\begin{eqnarray}}
\newcommand{\EA}{\end{eqnarray}}

\title{A variational study of bound states \\
                         in the Higgs model}
\author{Fabio Siringo}
\address{Dipartimento di Fisica e Astronomia,
Universit\`a di Catania \\
Corso Italia 57, I 95129 Catania, Italy}
\date{\today}
\maketitle
\begin{abstract}
The possible existence of Higgs-Higgs bound states in the Higgs sector 
of the Standard Model is explored using the $|hh\rangle + |hhh\rangle$ 
variational ansatz of Di Leo and Darewych.  The resulting integral 
equations can be decoupled exactly, yielding a one-dimensional integral 
equation, solved numerically.  We thereby avoid the extra approximations 
employed by Di Leo and Darewych, and we find a qualitatively different 
mass renormalization.  Within the conventional scenario, where a 
not-too-large cutoff is invoked to avoid ``triviality,'' we find, as 
usual, an upperbound on the Higgs mass.  Bound-state solutions 
are only found in the very strong coupling regime, but at the same time
a relatively small physical mass is required as a consequence
of renormalization.
\end{abstract}


\section{Introduction}

The possible existence of bound states for the Higgs boson has been 
studied by several authors \cite{suzuki,5inrupp,rupp,dileo} with 
both perturbative and non-perturbative calculations. At present, there
is little agreement between the quantitative predictions of such 
calculations.  

A variational method, within the Hamiltonian formalism\cite{schiff},
has been used by
Di Leo and Darewych (DLD) \cite{dileo}.  However, because of the apparent 
complexity of the resulting integral equations, they resorted to additional 
approximations that are unsatisfactory when the coupling is strong.  In 
this paper we show that an exact decoupling of DLD's integral equations is 
possible by use of some symmetry properties.  This allows a considerable 
simplification of the problem.  For an $s$-wave solution we show in detail 
that the method gives rise to a one-dimensional integral equation that can 
be tackled numerically.

Mass renormalization (beyond normal ordering) plays a crucial role, although 
it is finite.  (The infinite mass renormalization in DLD turns out to be an 
artifact of their other approximations).  The physical mass is significantly 
reduced from its classical value in the strong-coupling regime.  Because of 
this mass-reduction effect we find that the occurrence of bound
states (i.e., solutions in the 2-particle sector with $E<2m$) is shifted
towards the very strong coupling regime, well beyond the reach of
any perturbative approximation. However a relatively small physical mass
($m\approx 0.1-0.5$ TeV) is required as a consequence of the same
renormalization effect.

Our paper is organized as follows: the variational method is described
in section II. A general prescription for the mass renormalization is then 
provided and discussed.  Section III deals with the delicate aspect of mass
renormalization through a variational one-particle trial state calculation
analogous to the two-particle trial state used in section II. The
existence of Higgs-Higgs bound states is discussed in section IV
and the results are compared with those of other authors.

Our discussion here uses the conventional framework that the Higgs 
theory is an effective theory with a large, but {\it finite} cutoff $\Lambda$ 
\cite{consoli}.

\section{Variational Method}

    Following DLD, our starting point is the Lagrangian
\BE
{\cal{L}}=-\half\partial_\mu h\partial^\mu h
-\half m_0^2 h^2-{1\over{3!}}\lambda v h^3-{1\over{4!}}\lambda h^4,
\label{lagrangian}
\EE
for a neutral, scalar Higgs field $h$.  For the present we regard $m_0$, 
$v$, and $\lambda$ as three independent bare parameters; only in 
section IV shall
we need to impose the constraint $m_0^2 = \frac{1}{3} \lambda
v^2$ that arises when this model is obtained from a spontaneously broken 
$\LP$ theory.  

   In the Schr\"odinger representation,
at $t=0$, the field is quantized in terms of creation and
annihilation operators 
\BE
h({\bf{x}})= \int \! {{d^3k}\over{ \sqrt{ (2\pi)^3 2\omega_{ {\bf{k}} } } }}
\, \left[ a_{ {\bf {k}}}
\exp(i{\bf{k}}\cdot {\bf{x}}) +
a^{\dagger}_ { {\bf {k}} } 
\exp( -i{\bf{k}}\cdot {\bf{x}}) \right]
\label{field}
\EE
satisfying the usual commutation relations
\BE
\left[ a_{\bf k}, {a^\dagger}_{\bf p}\right]=\delta^3({\bf k}-{\bf p}).
\label{commutation}
\EE
The energy of the single particle states is
\BE
\omega(\kk)=\omega_k=\sqrt{\kk^2+m^2}
\label{omega}
\EE
where $m$ is the physical mass which may differ from the classical mass
$m_0$.
The Hamiltonian $H$ is obtained from Eq. (\ref{lagrangian}) as
\BA
H=\int \! d^3 k\left(\omega_k-{{\delta m^2}\over {2\omega_k}}\right)
{a^\dagger_\kk} a_\kk-\int \! d^3 k \, {{\delta m^2}\over{4 \omega_k}}
\left(a^\dagger_\kk a^\dagger_{-\kk}+ a_{\kk} a_{-\kk}\right)
+\qquad\qquad\qquad&&\nonumber\\
+:\left[{{\lambda v}\over {3!}}\int \! d^3 x \, h^3+
{\lambda\over{4!}}\int \! d^3 x \, h^4\right]:&&
\label{hamiltonian}
\EA
where $\delta m^2=m^2-m_0^2$.  The Hamiltonian has been normal 
ordered with respect to the physical mass $m$.  

   The preceding discussion follows the conventions of Ref. \cite{dileo}, 
except that our $\lambda$ is a factor of 6 larger than theirs, since we use 
$\lambda/4!$ rather than $\lambda/4$.  

    The trial two-boson bound state considered by DLD 
\cite{dileo} is 
\BE
\vert\Psi_2\rangle=\int \! d^3 p \, 
B(\pp) a^\dagger_\pp a^\dagger_{-\pp}\vert 0
\rangle+\int \! d^3 p \, d^3 q \, d^3 k \, 
G(\pp,\qq,\kk) a^\dagger_\pp a^\dagger_\qq
a^\dagger_\kk \vert 0\rangle \delta^3 (\pp+\qq+\kk) ,
\label{trial}
\EE
with $\vert 0\rangle$ the vacuum annihilated by $a$.  We observe that 
the function $B(\pp)$ may be taken to be symmetric without any loss of 
generality; a general $B(\pp)$ could always be decomposed into an even 
and an odd part and the odd part would give no contribution to (\ref{trial}). 
This reflects the fact that a bound state of two identical bosons must be 
even under spatial inversion.  Similarly, we must have 
$G(\pp,\qq,\kk)=G(-\pp,-\qq,-\kk)$.  Furthermore, there is no loss of 
generality in assuming that $G(\pp,\kk,\qq)$ is invariant under any 
permutation of the three momenta $\pp$, $\kk$ and $\qq$.  (Of course, 
because of the momentum-conserving delta function, the function $G$ really 
involves only two {\it independent} momentum arguments.)

    The $B$ and $G$ functions are determined from the variational principle
\BE
\delta\langle\Psi_2\vert H-E\vert\Psi_2\rangle=0
\label{variational}
\EE
which provides two coupled eigenvalue equations:
\BE
{{\delta\langle\Psi_2\vert H-E\vert\Psi_2\rangle}\over
{\delta B^\star(\kk)}}=0
\label{eigen1a}
\EE
\BE
{{\delta\langle\Psi_2\vert H-E\vert\Psi_2\rangle}\over
{\delta G^\star(\pp,\qq,\kk) }}=0
\label{eigen1b}
\EE
Explicitly, these equations are:
\BA
\left(2\omega_k-{{\delta m^2}\over{\omega_k}}-E\right)B(\kk)+
{\lambda\over{64\pi^3 \omega_k}}\int \! d^3p \,{{B(\pp)}\over{\omega_p}}
\, + \qquad&&
\nonumber\\
{\mbox{}}+{{3\lambda v}\over {8 \pi^{3/2}}}\int \! d^3p \! \int \! d^3q \,
\delta^3(\kk+\pp+\qq)) {{1}\over{\sqrt{\omega_k \omega_p \omega_q}}}
G(\kk,\pp,\qq) &=&0
\label{eigen2a}
\EA
\BA
\left[\sum_{i=1}^3\left(\omega_i-{{\delta m^2}\over{2\omega_i}}\right)
-E\right]G(\kk_1,\kk_2,\kk_3)
+{{\lambda v}\over{(4!)\pi^{3/2}\sqrt{\omega_1\omega_2\omega_3}}}
\sum_{i=1}^3 B(\kk_i)
&+&\nonumber\\
{\mbox{}}+{\lambda\over{64\pi^3\sqrt{\omega_1\omega_2\omega_3 }}}
\sum_{i=1}^3\int \! d^3p \! \int \! d^3q \, \delta^3(\kk_i+\pp+\qq)
\left({{\omega_i}\over{\omega_p \omega_q}}\right)^{1/2} 
G(\kk_i,\pp,\qq) &=&0,
\label{eigen2b}
\EA
where $\omega_i\equiv \omega_{k_i}$.  These equations are equivalent to 
Eqs.~(8),(9) of Ref.\cite{dileo}, but their structure appears considerably 
simpler because we have taken advantage of the symmetry properties 
mentioned above.  We can achieve an exact decoupling of these equations 
by the following algebraic manipulations.  First, we introduce the auxiliary
function 
\BE
A(\kk)=\omega_k \int \! d^3p \! \int \! d^3q \, \delta^3(\kk+\pp+\qq) 
{{G(\kk,\pp,\qq)}\over{\sqrt{\omega_k\omega_p\omega_q}}}.
\label{A}
\EE
so that the first equation becomes
\BE
{{3\lambda v}\over{8 \pi^{3/2} \omega_k}} A(\kk)=\left(E-2\omega_k 
+{{\delta m^2}\over{\omega_k}}\right) B(\kk)-
{{\lambda}\over{64\pi^3\omega_k}}\int \! \frac{d^3p}{\omega_p} B(\pp)
\label{eigen3a}
\EE
Similarly, re-writing its last term in terms of $A(\kk_i)$, Eq. (\ref{eigen2b}) 
becomes
\BE
G(\kk_1,\kk_2,\kk_3)=\frac{-\lambda}{64 \pi^3} 
\frac{1}{\sqrt{\omega_1\omega_2\omega_3}} 
\frac{\sum_{i=1}^3\left[A(\kk_i)+(8\pi^{3/2}/3) v B(\kk_i)\right]}
{\left[ \sum_{i=1}^3 \left(\omega_i-{{\delta m^2}\over{2\omega_i}}\right)
-E \right]} .
\label{G2}
\EE
Note that the resulting form of $G$ manifestly respects the symmetry properties 
invoked earlier.  Inserting this equation back into Eq. (\ref{A}), the second 
eigenvalue equation Eq. (\ref{eigen2b}) is equivalent to
\BA
A(\kk)+(8 \pi^{3/2}/3) {{J(\kk)}\over{1+J(\kk)}} v B(\kk) + 
\qquad\qquad\qquad & & 
\nonumber\\
{\mbox{}} + {{2 \omega_k}\over{1+J(\kk)}} \int \! d^3p \! \int \! d^3q \, 
\delta^3(\kk+\pp+\qq) 
K(\kk,\pp,\qq) \left[A(\pp)+(8 \pi^{3/2}/3) v B(\pp)\right]=0& &
\label{eigen3b}
\EA
where the kernel $K(\kk,\pp,\qq)$ is wholly symmetric: 
\BE
K(\kk_1,\kk_2,\kk_3)={{\lambda}\over{64 \pi^3}}
{{1}\over{\omega_1\omega_2\omega_3\left[
\sum_{i=1}^3\left(\omega_i-{{\delta m^2}\over{2\omega_i}}\right)-E\right]}}
\label{kernel}
\EE
and
\BE
J(\kk)=\omega_k \int d^3p \, K(\kk,\pp,-\kk-\pp) 
\label{integral}
\EE
is a logarithmically divergent integral.  We shall regularize it with an 
energy cutoff, $\sqrt{{\bf p}^2 + m^2} < \Lambda$. 

The eigenvalue equations (\ref{eigen3a}), (\ref{eigen3b}) may now be
easily decoupled by replacing $A$ from Eq. (\ref{eigen3a}) into
Eq. (\ref{eigen3b}). In this way, 
we obtain the following integral equation for $B$
\BA
\left[E - 2\omega_k + {1\over{\omega_k}}\left(\delta m^2+\lambda v^2
{{J(\kk)}\over{1+J(\kk)}}\right) \right]B(\kk)=
{{\lambda}\over{64 \pi^3 \omega_k}} {{1+3J(\kk)}\over{1+J(\kk)}}
\int {{d^3p}\over{\omega_p}} B(\pp) \, -
\quad 
&&\nonumber\\
{\mbox{}}-{{2}\over{(1+J(\kk))}} \int \! d^3p \, K(\kk,\pp,-\kk-\pp) B(\pp)
\left(\delta m^2+\lambda v^2+\omega_p(E - 2\omega_p) \right).&&
\label{eqint1}
\EA

  For an $s$-wave $B$ function, the angular integration can
be performed analytically, yielding a one-dimensional integral
equation for $B$ which can be solved by numerical methods.  
There are two main conceptual problems that must first be dealt with 
(i) the mass renormalization parameter $\delta m^2$ needs to be determined, 
and (ii) the integral $J$ is logarithmically divergent and requires 
regularization, say with an energy cut-off $\Lambda$.  
This last point has to do with the physical interpretation of the
whole theory.  The current orthodox viewpoint is that the original 
$\lambda\Phi^4$ theory is only an effective theory, valid up to some finite 
energy scale $\Lambda$ that acts as a cutoff.  $\Lambda$ is then another 
parameter of the theory, in addition to $m_0$ and $\lambda$.  We shall 
adopt this viewpoint in this paper.  (For a heterodox viewpoint, see 
Ref. \cite{consoli}.)  

  Mass renormalization is crucial since the existence of bound states 
hinges on the comparison between the energy $E$ and the energy of two free 
bosons at rest, $2m$.  Any attractive self-interaction that tends to bind 
two particles will also give rise to a reduction of the physical free-particle 
mass compared to the classical mass $m_0$.  Thus it would not be legitimate to 
ignore mass renormalization and just impose $m=m_0$. 

   The form of the left-hand side of equation (\ref{eqint1}) suggests 
\cite{dileo} that the desirable mass renormalization is such that 
the combination 
\BE
\label{lefthnd}
\left(\delta m^2+\lambda v^2 {{J(k)}\over{1+J(k)}}\right) 
\EE
should vanish.  Since $m^2$ should not be $k$ dependent, we define 
\BE
\delta m^2=-\lambda v^2{{J(0)}\over{1+J(0)}}.
\label{deltam}
\EE
For an infinite cutoff, $J \to \infty$ and we would get 
\BE
m^2=m_0^2-\lambda v^2
\label{mass}
\EE
which is a {\it finite} mass renormalization.  [In DLD, due to their other 
approximations, the $1+J(0)$ denominator is absent in Eq. (\ref{deltam}), 
so that they found an infinite mass renormalization.]  

     In the next section, we will show that the above mass renormalization is 
justified by considering a variational calculation of a one-particle 
state. 

\section{ Single-particle mass renormalization}

The mass renormalization prescription (\ref{deltam}) may be recovered
in an analogous self-consistent variational procedure. In this case, 
the trial state $\vert\Psi_\kk\rangle$ for a single boson with mass $m$
and momentum $\kk$ is taken to be
\BE
\vert\Psi_\kk\rangle=C(\kk) a^\dagger_\kk\vert 0\rangle+
\int \! d^3 p \, D(\kk,\pp) a^\dagger_{\pp+\kk} a^\dagger_{-\pp}\vert 0
\rangle
\label{trial1}
\EE
where $D(\kk,\pp)=D(\kk,-\pp-\kk)$.
The variational principle now requires that
\BE
\delta\langle\Psi_\kk\vert H-E(\kk)\vert\Psi_\kk\rangle=0
\label{variational1}
\EE
which leads to the coupled eigenvalue equations
\BE
C(\kk)\left(\omega_\kk-{{\delta m^2}\over{2\omega_\kk}}-E(\kk)\right)
+{{\lambda v}\over{8\pi^{3/2}}} \int{{d^3 p D(\kk,\pp)}\over
{\sqrt{\omega_\kk \omega_\pp \omega_{\kk+\pp}}}}=0
\label{eigen4a}
\EE
\BA
D(\kk,\qq)\left[\omega_\qq+\omega_{\kk+\qq}-{{\delta m^2}\over{2}}
\left({1\over{\omega_\qq}}+{1\over{\omega_{\kk+\qq}}}\right)-E(\kk)\right]
+{{\lambda}\over{64 \pi^3}}\int{{d^3 p \, D(\kk,\pp)}\over
{\sqrt{\omega_\qq\omega_{\kk+\qq}\omega_\pp\omega_{\pp+\kk}}}} \, +&&
\nonumber\\
{\mbox{}}+{{\lambda v}\over{16\pi^{3/2}}}{{C(\kk)}\over{\sqrt{\omega_\kk
\omega_\qq \omega_{\kk+\qq}}}}=0&&.
\label{eigen4b}
\EA
Eliminating the integral we obtain
\BE
{{D(\kk,\qq)}\over{C(\kk)}}={{1}\over{8\pi^{3/2}v}}\sqrt{{\omega_\kk}\over
{\omega_\qq\omega_{\kk+\qq}}}
\left[{{\omega_\kk-{{\delta m^2}\over{2\omega_\kk}}
-{{\lambda v^2}\over{2\omega_\kk}}-E(\kk)}\over{\omega_\qq+\omega_{\kk+\qq}
-{{\delta m^2}\over{2}}\left({1\over{\omega_\qq}}+
{1\over{\omega_{\kk+\qq}}}\right)-E(\kk)}}\right] .
\label{ratio}
\EE
Substituting back in Eq. (\ref{eigen4a}) then for $C\not=0$ we find
\BE
E(\kk)=\omega_\kk-{1\over{2\omega_\kk}}\left[\delta m^2+\lambda v^2
{{J_0 (\kk)}\over{1+J_0 (\kk)}}\right]
\label{dispersion}
\EE
where
\BE
J_0(\kk)={{\lambda}\over{64 \pi^3}} \int^\Lambda d^3p \,  
{{1}\over{\omega_\pp\omega_{\kk+\pp}
\left[\omega_\pp+\omega_{\kk+\pp}-{{\delta m^2}\over 2}
\left({1\over{\omega_\pp}}+{1\over{\omega_{\kk+\pp}}}\right)-
E(\kk)\right]}} .
\label{integral0}
\EE
This has a similar structure and the same ultraviolet behaviour as the 
integral $J(\kk)$ of Eq. (\ref{integral}).  In principle, the presence of 
the single particle energy $E(\kk)$ in the denominator requires us to solve
Eq. (\ref{dispersion}) and Eq. (\ref{integral0}) self-consistently.  
However, when $\Lambda\gg m$ (as it should be)
\BE
J_0(\kk)\approx J(\kk)\approx {{\lambda}\over{128 \pi^3}}\int^\Lambda
{{d^3 k}\over{\vert\kk\vert^3}}\approx {{\lambda}\over{32\pi^2}}
\ln (\Lambda/m)+ \mbox{{\rm finite terms}} 
\label{logarithm}
\EE
In such a limit, we self-consistently obtain $E(\kk)=\omega_\kk$ 
provided we take 
\BE
\delta m^2=-\lambda v^2 {{J_0}\over{1+J_0}}
\label{deltam1}
\EE
to be compared to Eq. (\ref{deltam}). Once more for $\Lambda\gg m$
we recover the mass renormalization prescription (\ref{mass}).

Full consistency would also require that when $E(\kk)=\omega_\kk$ the
function $D$ should vanish. In fact, since $\omega_\kk$ is the energy of
a single particle state $\vert \kk\rangle=a^\dagger_\kk \vert 0\rangle$
with mass $m$, the state $\vert \Psi_\kk\rangle$ can
have a mass $m$ and a Lorentz-covariant dispersion relation  only 
when $\vert\Psi_\kk\rangle=\vert \kk\rangle$. This requirement is not
entirely trivial: the mass renormalization prescription (\ref{deltam1})
is just what we need in order to guarantee that the single particle state
$\vert \kk\rangle$ effectively is the lower energy one-particle
state for the full interacting Hamiltonian (\ref{hamiltonian}).
Inserting Eq. (\ref{dispersion}) and (\ref{deltam1}) into Eq. (\ref{ratio}),
we find that the ratio $D/C\to 0$ for $\Lambda\to\infty$ as we expected.
In other words that means we find a quantum-field renormalization constant
$Z=1$.

It is instructive examining the same result from the point of view of
standard perturbation theory. If there were some ``rule" forbidding the 
existence of states with more than two particles in the Fock space, then the
variational trial state (\ref{trial1}) would lead to an exact result.
The same result should be achievable by perturbation theory provided
that we sum all Feynman diagrams whose intermediate states do not contain 
more than two particles. However, in the self-consistent
variational procedure the mass of the single
particle state $\vert \kk\rangle$ is supposed to be the true
physical mass $m\not=m_0$. In the language of perturbation theory
this is equivalent to associate a renormalized propagator with each
internal line of Feynman diagrams. The easiest way to do that is by
re-writing the Lagrangian (\ref{lagrangian}) as
\BE
{\cal {L}}={\cal {L}}_0+{\cal {L}}_1+{\cal{L}}_2
\label{L}
\EE
where ${\cal{L}}_0$ is the zeroth-order non-interacting part
\BE
{\cal{L}}_0=-\half\partial_\mu h_R\partial^\mu h_R-\half m^2 h_R^2
\label{L0}
\EE
${\cal{L}}_1$ is an interaction part contributing at tree level
\BE
{\cal{L}}_1=-\half(Z-1)\partial_\mu h_R\partial^\mu h_R-\half (Z-1)m^2 h_R^2
+\half Z\delta m^2 h_R^2
\label{L1}
\EE
and ${\cal{L}}_2$ is the interaction
\BE
{\cal{L}}_2=
-{1\over{3!}}\lambda vZ^{3/2} h_R^3-{1\over{4!}}Z^2\lambda h_R^4
\label{L2}
\EE
Here $h_R$ is a renormalized field $h_R=h/\sqrt{Z}$.
Imposing that the renormalized propagator has a pole at $p^2=-m^2$ with
unit residue gives two conditions\cite{book}:
\BE
Z\delta m^2=-\Pi^\star_{loop}(-m^2)
\label{cond1}
\EE
and
\BE
Z=1+{{d}\over{dp^2}} \Pi^\star_{loop}(p^2)\vert_{-m^2}
\label{cond2}
\EE
where $p^2=\pp^2-\omega_\pp^2$, and
$i (2\pi)^4\Pi^\star_{loop}(p^2)$ is the sum of all one-particle-irreducible
diagrams containing loops. Such diagrams can only arise from the lagrangian
part ${\cal{L}}_2$. In our reduced Fock space
the only diagrams contributing are the bubble-chain diagrams
reported in Fig.1. These are all naively divergent but, if a regularization
prescription allows their resummation, 
their infinite sum yields
a finite contribution
\BE
\Pi^\star_{loop}(p^2)=
\Pi^\star_{1-loop}+\Pi^\star_{1-loop}(p^2)\cdot\left[
-{1\over 2} Z^2\lambda I(p^2)\right]+
\Pi^\star_{1-loop}(p^2)\cdot\left[-{1\over 2} Z^2\lambda I(p^2)\right]^2+
\dots
\label{series}
\EE
where the one-loop term is
\BE
\Pi^\star_{1-loop} (p^2)={1\over 2} (Z^3\lambda^2 v^2) I(p^2)
\label{1loop}
\EE
and $I(p^2)$ is the divergent integral
\BE
I(p^2)=\int{{d^4 q}\over{i(2\pi)^4}} {{1}\over{
\left(q^2+m^2-i\epsilon\right)\left( (p+q)^2+m^2-i\epsilon\right)}}
\label{integral1}
\EE
The infinite series (\ref{series}) can be exactly summed up
\BE
\Pi^\star_{loop}(p^2)=(Z\lambda v^2)\left[{{ {1\over 2}\lambda Z^2 I(p^2)}
\over{1+{1\over 2}\lambda Z^2 I(p^2)}}\right]
\label{allloops}
\EE
In order to mantain Lorentz-covariance, the integral (\ref{integral1})
can be evaluated by dimensional regularization. By use of the Feynman
formula and Wick rotation, in a $d$ dimensional Euclidean space the
integral reads
\BE
I(p^2)={{\pi^{d/2}}\over{(2\pi)^4}} \Gamma(2-d/2)  \int_0^1
\left[m^2+p^2 x(1-x)\right]^{ {d\over 2}-2} dx
\label{dintegral}
\EE
For $d=4+\epsilon$ and $\epsilon\to 0$ we obtain
\BE
I(p^2)=-{1\over {8\pi^2\epsilon}}-{1\over {16\pi^2}}
\left[\gamma+\ln\pi+\int_0^1 dx\ln\left(m^2+p^2 x(1-x)\right)\right]
+{\cal{O}}(\epsilon)
\label{epsintegral}
\EE
Insertion in Eq. (\ref{allloops}) gives
\BE
{{d}\over{dp^2}} \Pi^\star_{loop}(p^2)\vert_{-m^2}\sim\epsilon^2
\EE
and from Eq. (\ref{cond2}) then
\BE
Z=1+{\cal {O}} (\epsilon^2)
\label{Z}
\EE
while from Eq. (\ref{cond1}) the mass renormalization parameter reads
\BE
\delta m^2=-\lambda v^2\left[{{\lambda/(16\pi^2 \epsilon)}
\over{\lambda/(16\pi^2 \epsilon)-1}}\right]
\label{masseps}
\EE
Thus, in the physical $d=4$ space, we recover for $\epsilon\to 0$
the mass renormalization prescription of Eq. (\ref{mass}) and $Z=1$.

\section{ Search for Higgs-Higgs bound states}

In the previous sections we described the variational method and
discussed its internal consistency from a general point of view.  
So far the three parameters $m_0^2$, $v$ and $\lambda$ have been viewed 
as independent and we have not specialized to any particular physical
problem.  We now wish to use the method described in section II to search for 
Higgs-Higgs bound states in the Higgs sector of the electroweak theory.  

The scalar sector of the electroweak theory has the form:
\BE
{\cal{L}}=-\half\partial_\mu \Phi\partial^\mu \Phi 
-{{\lambda}\over{4!}}(\Phi^2 - v^2)^2 + {\rm const.}
\EE
Defining the `Higgs' field $h$ by $\Phi=v+h$, we obtain the original 
Lagrangian (\ref{lagrangian}) with the parameters related by 
\BE
\label{clmass}
m^2_0={1\over 3} \lambda v^2 .
\EE
It can be shown that the same relation holds for $v$ being the minimum
of the Gaussian effective potential\cite{stevenson}.
In the standard interpretation, we also have a large but finite cutoff
$\Lambda$.  The theory is approximately Lorentz invariant for energies 
small compared to a finite energy cut-off $\Lambda$.  
The vacuum value $v$ is fixed empirically in terms of the Fermi constant 
$G_F$ 
\BE
2 v^2={{\sqrt{2}}\over{G_F}}
\label{fermi}
\EE
Thus the bare mass $m_0$ is proportional to the square root of
the coupling $\lambda$. The physical model is entirely described by
two independent energy scales: the energy cut-off $\Lambda$ and the
bare mass $m_0$ which also fixes the coupling strength through
Eq. (\ref{clmass}). 

   For a large cut-off $\Lambda$, according to Eq. (\ref{mass}), we expect 
a mass correction $\delta m^2\approx -\lambda v^2$ which overcomes the 
tree-level mass (\ref{clmass}).  Since $m^2$ is positive definite we expect 
that something should prevent it from becoming negative.
In fact the general discussion of section II must be modified when
the physical mass $m$ approaches zero. At the point $m\to 0$,
$\Lambda\to\infty$ the integral $J(0)$ is not analytical, and some
extra care is required in handling the two limits. As previously
discussed $J(0)$ diverges logarithmically according to
Eq. (\ref{logarithm}) for any finite $m$, while it vanishes linearly
for $m\to 0$ at any fixed cut-off $\Lambda$. Thus for any large but
finite $\Lambda$ the coupled equations (\ref{integral}) and (\ref{deltam})
must be solved together yielding a real cut-off dependent mass $m(\Lambda)$.
Of course at any finite cut-off such equations mantain a $k$-dependence
since the theory is not Lorentz invariant. We define $m$ as the $\kk=0$
value corresponding to the energy required to create a boson at rest.
For $\kk=0$ a generic scattering solution of
the integral equation (\ref{eqint1}) has $E=2m$ where $m$ is determined
from Eq. (\ref{deltam}) by insertion of Eq. (\ref{clmass})
\BE
m^2=m^2_0\left[{{1-2J(0)}\over{1+J(0)}}\right]
\label{self1}
\EE
while $J(0)$ is the integral (\ref{integral}) of the kernel (\ref{kernel})
evaluated at $\kk_1=0$ and at $E=2m$:
\BE
J(0)={{\lambda}\over{32\pi^2}}\int_1^g
{{\sqrt{x^2-1}}\over{x^2-\alpha x
-\beta}}dx
\label{self2}
\EE
where $g=\Lambda/m$, $\alpha=(3-m^2_0/m^2)/4$ and $\beta=(1-m^2_0/m^2)/2$.

The two coupled equations (\ref{self1}) and (\ref{self2}) give the
physical mass $m$. The integral in Eq. (\ref{self2}) can be evaluated 
analytically:  
\BE
J(0)={{3 m^2_0}\over{32\pi^2 v^2}}f(g) ,
\label{self3}
\EE
where
\BA
f(g)=\ln\left(g+\sqrt{g^2-1}\right)&+&{1\over 2}
\sum_{\pm}{{\sqrt{\gamma_\pm^2-1}}
\over{\pm\sqrt{\Delta}}}
\ln\left(1-g\gamma_\pm+\sqrt{g^2-1}\sqrt{\gamma^2_\pm-1}\right)\nonumber\\
&-&{1\over 2}\sum_{\pm}{{\sqrt{\gamma_\pm^2-1}}
\over{\pm\sqrt{\Delta}}}
\ln\left(1-g\gamma_\pm-\sqrt{g^2-1}\sqrt{\gamma^2_\pm-1}\right) ,
\label{numint}
\EA
and 
$\gamma_\pm=(\alpha\pm\sqrt{\Delta})/2$ and $\Delta=\alpha^2+4\beta$.

A numerical solution for $m$ as a function of the coupling parameter
$m_0$ is reported in Fig. 2 for several cut-off values.
In the weak-coupling limit, for $m_0$ smaller than $0.3$ TeV, we recover
the perturbative solution $m\approx m_0$ which holds up to a quite huge
$\Lambda$. That is equivalent to neglecting $J(0)$ altogether in 
Eq. (\ref{self1}).  For larger couplings the solution deviates from the 
perturbative regime and the physical mass $m$ is heavily reduced and 
strongly dependent on the cut-off choice. Notably in the range 
$1$ TeV$~<m_0<2$ TeV, where several authors find Higgs-Higgs 
bound states, the physical mass is spread over a large energy range going 
from $m=m_0$ for $\Lambda=m_0$ to $m\approx m_0/100$ for $\Lambda=35$ TeV. 
However the upper bound $m=m_0$ is only reached for an unphysical cut-off 
equal to the mass, which makes the integral $J$ vanish. A cross-over is 
observed at a critical $\Lambda_c= 3.05$ TeV from a monotonic increasing 
behaviour of $m$ versus $m_0$, to a non-monotonic beahviour with $m$ 
rising to a maximum and then decreasing for larger $m_0$. The maximum value
of $m$ never exceeds the critical value $m_c=1.1$ TeV (which is reached at
a coupling $m_0=4$ TeV) for any choice of coupling and cut-off.  Thus, 
for any $\Lambda> 3.05$ TeV and any $m_0$, we always find a physical mass 
$m< 1.1$ TeV.  

   Moreover, in the strong coupling regime the physical mass becomes very
small: in such strong-coupling limit a simple analytical solution
may be found by requiring that $m\ll m_0$ and expanding both
equations (\ref{self1}) and (\ref{self2}) in powers of $m^2/m_0^2$.
By use of the analytical expression (\ref{numint}), the integral
(\ref{self2}) may be written as
\BE
J(0)={{3 m \Lambda}\over{8\pi^2 v^2}}+{\cal {O}}(m^2/m_0^2)
\label{self2bis}
\EE
which to first order in $m$ does not depend on the coupling $m_0$.
Eq. (\ref{self1}) may be inverted and to the same order in $m^2$ reads
\BE
J(0)={1\over 2} +{\cal{O}}(m^2/m_0^2)
\label{self1bis}
\EE
Eliminating $J(0)$ yields
\BE
m \approx {4\over 3} \pi^2 {{v^2}\over \Lambda}
\label{approxm}
\EE
which is consistent with our assumption that $m\ll m_0$ provided that
$m_0\gg 4\pi^2 v^2/(3\Lambda)$ or explicitly $m_0\cdot\Lambda\gg 0.8$ TeV$^2$.
When such condition is satisfied the physical mass does not depend on
the coupling and tends to a finite limit proportional to $\Lambda^{-1}$.
This behaviour is evident in Fig. 2.

For completeness we should mention that whenever $m_0>\Lambda$ a second
larger solution for the physical mass $m$
comes out from the coupled equations (\ref{self1}), (\ref{self2}), but 
such solutions probably have no physical meaning.  We remark that the 
range $m_0>\Lambda$ could be of physical interest since the cut-off 
$\Lambda$ should be compared to the physical mass $m$ which is generally 
significantly smaller than the coupling parameter $m_0$.

Returning to the bound-state problem, let us insert the
numerical solution of the coupled
equations  (\ref{self1}), (\ref{self3}) into the integral equation
(\ref{eqint1}).  Neglecting the slight $\kk$-dependence of 
$J(\kk)\approx J(0)$ (which is the only approximation we are making apart 
from the choice of the trial state) we obtain the following integral equation
\BE
(2\omega_k-E) B(\kk)=-\int \! d^3p {\cal {K}}(\kk,\pp,-\kk-\pp) B(\pp) ,
\label{eqint2}
\EE
where the kernel ${\cal{K}}$ is defined as 
\BA
{\cal {K}}(\kk,\pp,\qq)={1\over{64\pi^3\omega_k}}\left({{2m_0^2+m^2}\over
{v^2}}\right)\times\qquad\qquad\qquad\qquad\qquad\qquad
\qquad\qquad\qquad\qquad&&\nonumber\\
\times\left\{\left({{5m^2_0-2m^2}\over{2m_0^2+m^2}}\right)
{1\over{\omega_p}}- {2\over{\omega_p\omega_q}}
{{m^2+2m_0^2+\omega_p(E-2\omega_p)}\over
{\left[\omega_k+\omega_p+\omega_q+{{m_0^2-m^2}\over 2}
\left({1\over{\omega_k}}+{1\over{\omega_p}}+{1\over{\omega_q}}\right)
-E\right]}}\right\}&& .
\label{kernel2}
\EA
We notice that the second term inside the brackets
contains an attractive part plus an
energy dependent part proportional to $(E-2\omega_p)$ which is
always repulsive if $E<2m$, and thus weakens the bonding of any bound
state. It is instructive to see how the approximate integral equation
of DLD \cite{dileo} can be recovered from our almost exact treatment
of the variational method. That can be done by taking the perturbative
limit ($J(0)=0$, $m=m_0$) and by neglecting the energy dependent
repulsive part proportional to $(E-2\omega_p)$ in the second term of
the kernel according to their approximation $E\approx 2\omega_p$.
Moreover, since
they also take $E\approx \omega_p+\omega_k\approx\omega_q+\omega_k$,
the denominator of the second term is approximated as
\BE
{1\over{\omega_k+\omega_p+\omega_q-E}}\approx{1\over{\omega_p}}\approx
{1\over{\omega_q}}\approx{1\over 2}
\left({1\over{\omega_p}}+{1\over{\omega_q}}\right)
\label{mutilation}
\EE
yielding the approximation
\BE
{\cal {K}}(\kk,\pp,\qq)\approx {{\lambda }\over{64 \pi^3}}
{{3 m^2}\over{\omega_k\omega_p}}
\left[{1\over{3m^2}}-{1\over{\omega_q^2}}-{1\over{\omega_q\omega_p}}\right]
\label{approx}
\EE
which is Eq. (16) of Ref.\cite{dileo}.  This approximation requires 
$m \approx m_0$, which is valid only in the region $m_0<0.3$ TeV, according 
to our numerical results in Fig.2.  DLD found bound-state solutions with the 
kernel (\ref{approx}), but only for larger couplings $m_0>0.9$ TeV; this is 
well beyond the perturbative regime \cite{dileo,10indileo,11indileo}, and well 
beyond the region of validity of their approximations.

   To address the problem of the existence of bound-states beyond the 
perturbative regime, we must use the full integral equation (\ref{eqint2}). 
The most interesting case is an $s$-wave solution, and in that case the 
integration over angles may be carried out exactly yielding a one-dimensional 
integral equation.  We change the integration variables according to
\BE
\int \! d^3p \ldots = 
{{2\pi}\over{\sqrt{\omega_k^2-m^2}}}\int_m^\infty \omega_p
d\omega_p\int_{\omega_-}^{\omega_+} \omega_q d\omega_q \ldots ,
\label{variables}
\EE
where
\BE
\omega_{\pm}=\sqrt{\omega_k^2+\omega_p^2-m^2\pm 2\sqrt{(\omega_k^2-m^2)
(\omega_p^2-m^2)}} .
\label{omegapm}
\EE
Then integrating over $\omega_q$ gives
\BE
(2\omega_k-E)B(\kk)=\int_m^\Lambda{\cal {F}}
(\omega_k,\omega_p)B(\pp)d\omega_p ,
\label{eqint3}
\EE
with 
\BA
{\cal {F}} (\omega_k,\omega_p)=-{{2m_0^2+m^2}\over{16\pi^2 v^2\omega_k}}
\left\{
\left({{5m_0^2-2m^2}\over{2m_0^2+m^2}}\right) \sqrt{\omega_p^2-m^2}
\right.+\qquad\qquad\qquad\qquad\qquad\qquad\qquad&&\nonumber\\
-\left.
{{2m_0^2+m^2+\omega_p(E-2\omega_p)}\over{\sqrt{\omega_k^2-m^2}}}
\left[
{{\Omega_+}\over{\Omega_+-\Omega_-}}
\ln\left({{\omega_+-\Omega_+}\over{\omega_- -\Omega_+}}\right)
-{{\Omega_-}\over{\Omega_+-\Omega_-}}
\ln\left({{\omega_+-\Omega_-}\over{\omega_- -\Omega_-}}\right)
\right]
\right\}&&
\label{kernel3}
\EA
where $\Omega_\pm$ are the poles of the kernel ${\cal{K}}$ in
the variable $\omega_q$
\BA
\Omega_\pm&=&{{E-\omega_k-\omega_p}\over 2}-{{m_0^2-m^2}\over 4}
\left({1\over{\omega_k}}+{1\over{\omega_p}}\right)+
\nonumber\\
&&\nonumber\\
&\pm&{1\over 2}\sqrt{
\left[E-\omega_k-\omega_p-{{m_0^2-m^2}\over 2}
\left({1\over{\omega_k}}+{1\over{\omega_p}}\right)\right]^2
-2m_0^2+2m^2} \, .
\label{bigomega}
\EA
The one-dimensional integral equation (\ref{eqint3}) may be solved
numerically by standard matrix techniques.  (One must first choose the 
parameters $m_0$ and $\Lambda$, and find the corresponding physical mass 
$m$ from the coupled equations (\ref{self1}), (\ref{self3}).)
We can describe three different scenarios:

i) $\Lambda<\Lambda_c=3.05$ TeV (small cut-off). Since $m$ is a
monotonic increasing function of the coupling strength $m_0$, both
$m$ and $m_0$ must be small compared to $\Lambda$. In this weak-coupling
limit there are no bound-state solution and the lower eigenvalue of
Eq.(\ref{eqint3})  is the free particle energy $E=2m$.

ii) $\Lambda>\Lambda_c$, $m_0<2$ TeV (large cut-off and moderately
strong coupling). Beyond the critical value $\Lambda_c=3.05$ TeV the
Higgs mass $m$ is not a monotonic increasing function of the coupling
strength $m_0$ (see Fig.2). This strength can be large since $m$ is
bounded and never comparable to $\Lambda$. Beyond the weak-coupling
limit, where there are no bound-state solutions, an
intermediate range can be described for $m_0\approx 1-2$ TeV.
In this range non-perturbative effects are evident since $m$ is
a {\it decreasing} function of the  bare mass $m_0$ (as shown in Fig.2).
In this range bound state solutions have been found by several
authors\cite{suzuki,5inrupp,rupp,dileo}. However the strong reduction
of the physical mass $m$, in comparison with the bare mass $m_0$, rules
out the existence of bound states with $E<2m$ in this regime.

iii) $\Lambda>\Lambda_c$, $m_0>2$ TeV (very strong coupling). For
very large $m_0$ the renormalized Higgs mass $m$ saturates at the finite
value given by Eq.(\ref{approxm}) (see also Fig.2).
A further increase of the coupling strength allows the occurrence of
bound-state solutions whose precise onset depends on the chosen energy
cut-off $\Lambda$. In Fig.3 we show the binding energy $(E-2m)$, in
units of $2m$, for a typical cut-off $\Lambda=4$ TeV,  just above
$\Lambda_c$. An eigenvalue $E$ smaller than $2m$ appears at $m_0=2.35$
TeV, and the binding energy reaches the bootstrap point
$(E-2m)/(2m)=-0.5$ at $m_0=3.5$ TeV.

Despite the huge couplings required for binding, the corresponding
physical mass is relatively small compared to that found in previous
works\cite{suzuki,5inrupp,rupp,dileo}. Fig.4 reports the binding energy
versus $m$ (physical mass) for $\Lambda=4$ TeV. The onset of the
bound-state solution is at $m=519$ GeV, while the mass bootstrap
point is reached at $m=386$ GeV. We notice that the binding energy
now {\it increases} with decreasing $m$. Moreover, according to
Eq.(\ref{approxm}), an even smaller mass is required for larger choices
of the energy cut-off $\Lambda$.

At the light of our study a Higgs-Higgs bound state would be
conceivable for $m\approx 100-500$ GeV provided that the coupling
is very strong $m_0\approx 2$ TeV. We stress the role played by
mass renormalization in determining both, the shift of bound states
towards higher coupling strengths, and the corresponding reduction
of the physical mass required for bonding.
Most of the previous
calculations should  be revised at the light of the present result
in order to establish if mass renormalization has been correctly
addressed. We just mention Rupp's\cite{rupp} Bethe-Salpeter
approach where the chosen subtraction point gives $m=m_0$ at any coupling.

\acknowledgements

I thank Maurizio Consoli and Paul Stevenson for their 
generous assistance in this research.

\begin{figure}
\caption{Bubble-chain diagrams contributing to $\Pi^\star_{loop}$
in Eq.(\ref{series})}
\end{figure}

\begin{figure}
\caption{The physical Higgs mass $m$ versus the bare mass $m_0$ (which
fixes the coupling strength), for several choices of the energy cut-off
$\Lambda$=2.8, 3.2, 3.6, 4.0, 4.4, 4.8, 5.2, 5.6, 6.0, 6.4, 6.8 and
7.2 TeV.
Data are also reported for the cross-over point from monotonic
to non-monotonic behaviour occurring at $\Lambda_c$=3.05 TeV.
The dashed line represents the $m=m_0$ approximation which only holds in the
weak-coupling regime $m_0<0.3$ TeV.}
\end{figure}

\begin{figure}
\caption{Binding energy $E-2m$ in units of $2m$ versus bare mass
$m_0$ (coupling strength) for a cut-off $\Lambda=4$ TeV.}
\end{figure}

\begin{figure}
\caption{Binding energy $E-2m$ in units of $2m$ versus physical Higgs
mass $m$ for a cut-off $\Lambda=4$ TeV.}
\end{figure}

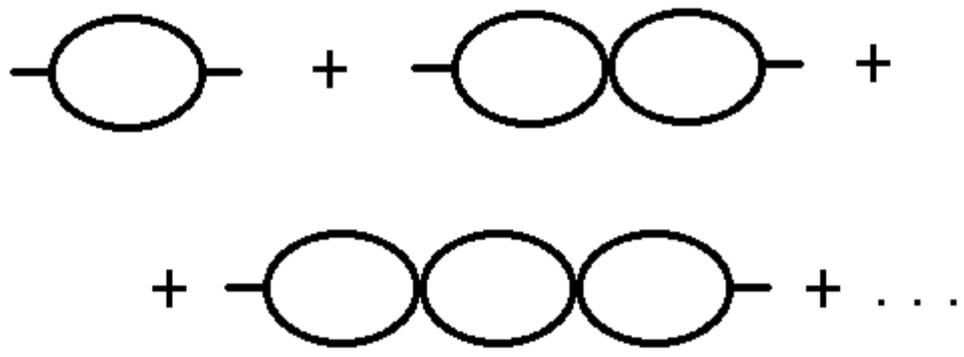
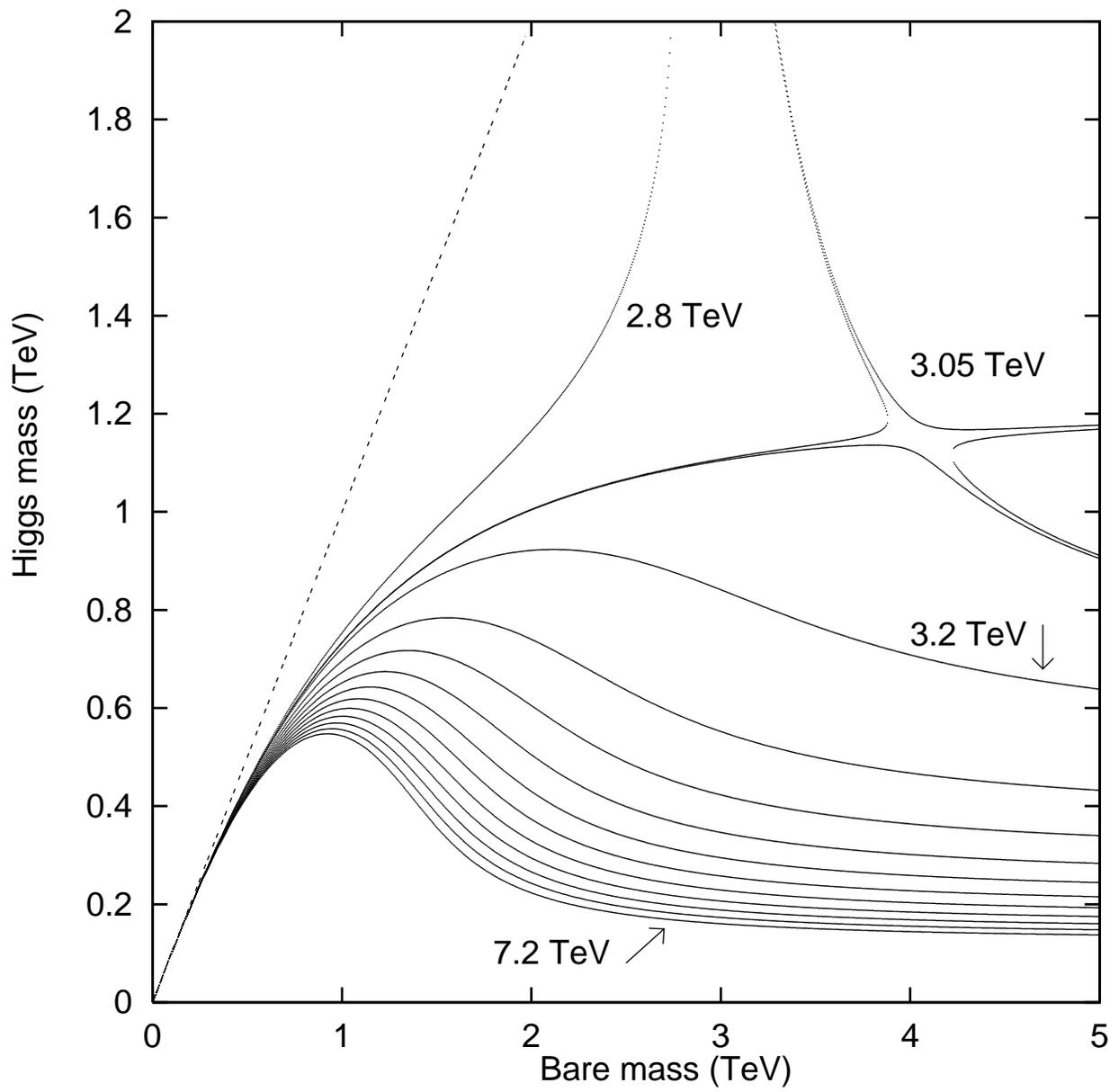
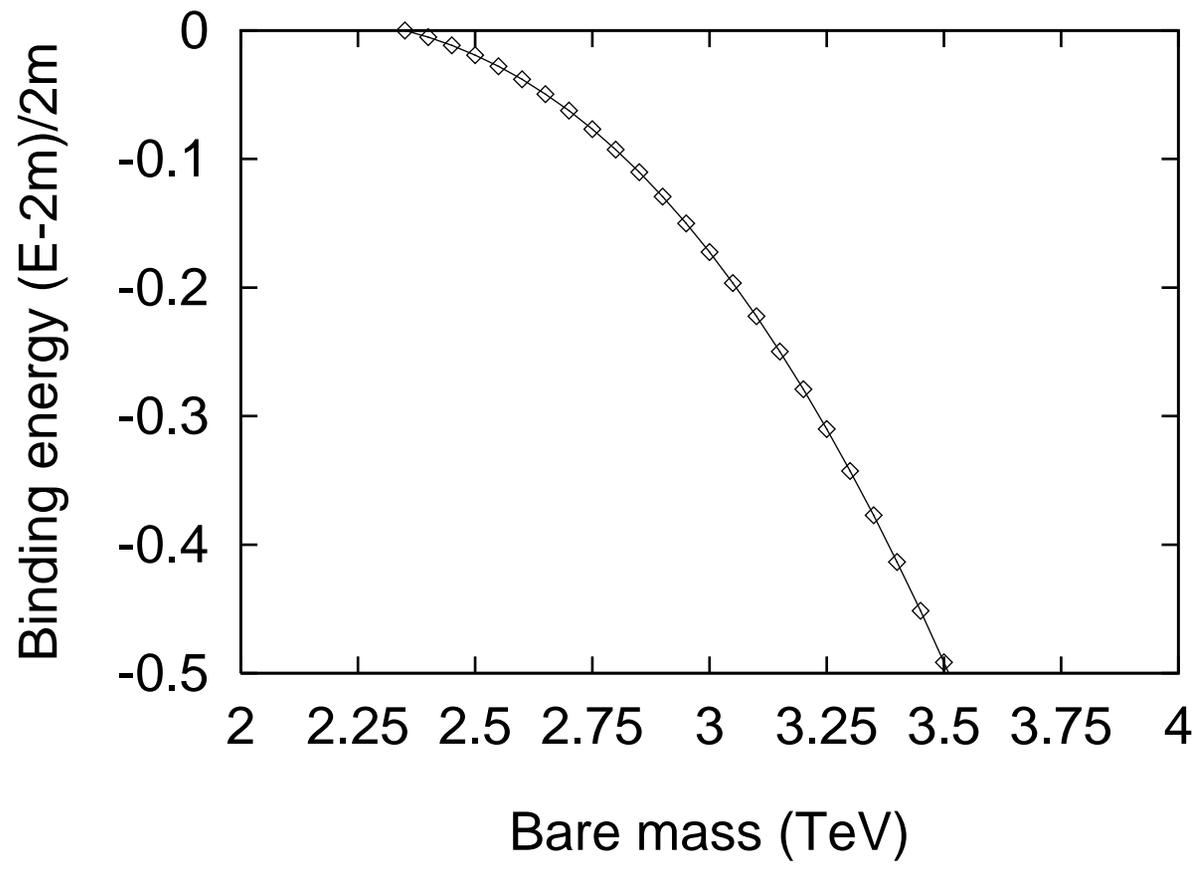
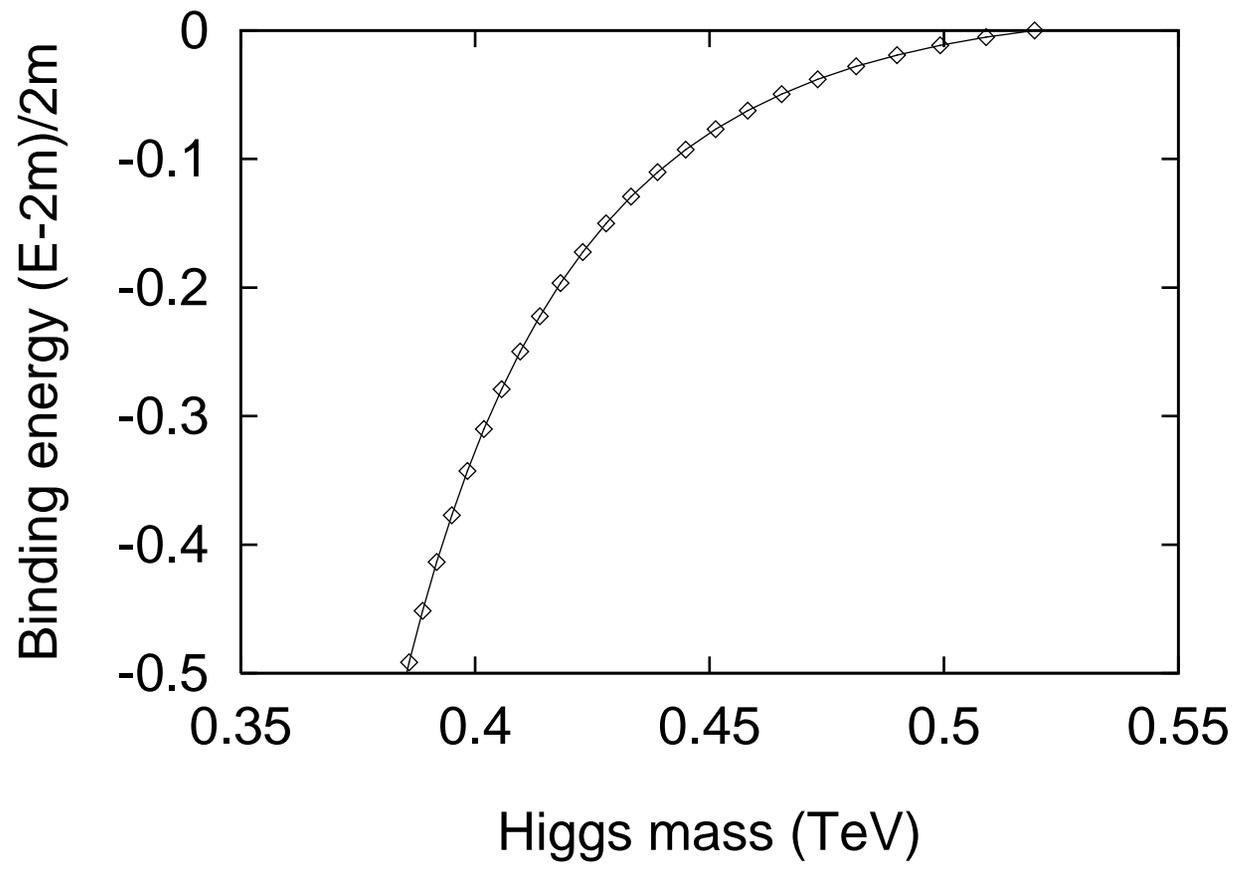
\end{document}